\documentclass[11pt,english]{article}
\usepackage[T1]{fontenc}
\usepackage[latin1]{inputenc}
\usepackage{babel}
\usepackage{graphics}

\makeatletter

\providecommand{\LyX}{L\kern-.1667em\lower.25em\hbox{Y}\kern-.125emX\@}

\usepackage{moriond,epsfig}

\def\be{\begin{equation}}
\def\ee{\end{equation}}
\def\bea{\begin{eqnarray}}
\def\eea{\end{eqnarray}}

\begin{document}
\vspace*{4cm}
\title{PERTURBATIONS GROWTH AND BIAS DURING ACCELERATION \footnote{To appear in the proceedings of the XXXVIIth Rencontres de Moriond, "The Cosmological Model", Les Arcs, France, March 2002.}}

\author{LUCA AMENDOLA AND DOMENICO TOCCHINI-VALENTINI}

\address{Osservatorio Astronomico di Roma, \\
 Viale Frascati 33, 00040 Monte Porzio Catone (Roma), Italy}

\maketitle\abstracts{
In most models of dark energy the structure formation stops when the accelerated expansion begins. In contrast, we show that the coupling of dark energy to dark matter may induce the growth of perturbations even in the accelerated regime. In particular, we show that this occurs in the models proposed to solve the cosmic coincidence problem, in which the ratio of dark energy to dark matter is constant. Moreover, if the dark energy couples only to dark matter and not to baryons, as requested by the constraints imposed by local gravity measurements, the baryon fluctuations develop a constant, scale-independent, large-scale bias which is in principle directly observable. }

As it has been shown \cite{wet95,ame}, an epoch of acceleration \cite{rie}
with a constant non-vanishing \( \Omega _{c} \) can be realized by
coupling dark matter to dark energy. In fact, a dark energy scalar
field \( \phi  \) governed by an exponential potential linearly coupled
to dark matter yields, in a certain region of the parameter space,
an accelerated expansion with a constant ratio \( \Omega _{c}/\Omega _{\phi } \)
and a constant parameter of state \( w_{\phi } \), referred to as
a stationary accelerated era. Similar models have been discussed elsewhere
\cite{pav}. We showed \cite{amtoc2} that in fact the conditions
of constant \( \Omega _{\phi } \) and \( w_{\phi } \) uniquely determine
the potential and the coupling of the dark energy field. The main
motivation to consider a stationary dynamics is that it would solve
the cosmic coincidence problem of the near equivalence at the present
of the dark energy and dark matter densities, since they scale identically
with time. Further theoretical motivations for coupled dark energy
have been put forward \cite{gasp,wet95,ame}.

Consider three components, a scalar field \( \phi  \), baryons and
CDM described by the energy-momentum tensors \( T_{\mu \nu (\phi )}, \)
\( T_{\mu \nu (b)} \)and \( T_{\mu \nu (c)} \), respectively. General
covariance requires the conservation of their sum, so that it is possible
to consider a coupling such that, for instance,\begin{eqnarray*}
T_{\nu (\phi );\mu }^{\mu } & = & \sqrt{2/3}\kappa \beta T_{(c)}\phi _{;\nu }\\
T_{\nu (c);\mu }^{\mu } & = & -\sqrt{2/3}\kappa \beta T_{(c)}\phi _{;\nu },
\end{eqnarray*}
 where \( \kappa ^{2}=8\pi G \), while the baryons are assumed uncoupled
\cite{dam,ame3}, \( T_{\nu (b);\mu }^{\mu }=0 \) because local gravity
constraints indicate a baryon coupling \( \beta _{b}<0.01 \) \cite{wet95,ame,dam96}.
Let us derive the background equations in the flat conformal FRW metric.
The equations for this model have been described in another work \cite{amtoc},
in which a similar model (but with a variable coupling) was studied.
We restrict ourselves to the case in which radiation has already redshifted
away. For the scalar field we assume a potential \( U(\phi )=U_{0}e^{-\sqrt{2/3}\mu \kappa \phi } \)
. The coupling \( \beta  \) can be seen as the relative strength
of the dark matter-dark energy interaction with respect to the gravitational
force. The only parameters of our model are \( \beta  \) and \( \mu  \)
(the constant \( U_{0} \) can always be rescaled away by a redefinition
of \( \phi  \)). 
The system is best studied in the new variables \cite{cop,ame3} \( x=\kappa \phi '/\sqrt{6},\quad y=\frac{\kappa a}{H}\sqrt{U/3}, \)
and \( v=\frac{\kappa a}{H}\sqrt{\rho _{b}/3} \), where the prime
stands for derivations respect to \( \log a \). The CDM energy density
parameter is obviously \( \Omega _{c}=1-x^{2}-y^{2}-v^{2} \) while
we also have \( \Omega _{\phi }=x^{2}+y^{2}, \) and \( \Omega _{b}=v^{2} \).
The system is subject to the condition \( x^{2}+y^{2}+v^{2}\leq 1 \).
The critical points of the resulting dynamical system are listed in
Tab. 1. We denoted with \( w_{e}=1+p_{tot}/\rho _{tot}=1+x^{2}-y^{2} \)
the total parameter of state. On all critical points the scale factor
expansion is given by \( a\sim \tau ^{p/1-p}=t^{p} \), where \( p=2/(3w_{e}) \),
while each component scales as \( a^{-3w_{e}} \). In the table we
also denoted \( g\equiv 4\beta ^{2}+4\beta \mu +18 \), and we used
the subscripts \( b,c \) to denote the existence of baryons or matter,
respectively, beside dark energy.  

\begin{table*}

\caption{Critical points.}
\begin{center}
\begin{tabular}{|ccccccc|}
\hline 
Point &
 \( x \)&
 \( y \)&
 \( v \)&
 \( \Omega _{\phi } \)&
 \( p \)&
\( w_{e} \)\\
\hline 
\( a \)&
 \( -\frac{\mu }{3} \)&
 \( \sqrt{1-\frac{\mu ^{2}}{9}} \)&
 0 &
 1 &
 \( \frac{3}{\mu ^{2}} \)&
\( \frac{2\mu ^{2}}{9} \)\\
 \( b_{c} \)&
 \( -\frac{3}{2\left( \mu +\beta \right) } \)&
 \( \frac{\sqrt{g-9}}{2\left| \mu +\beta \right| } \)&
 \( 0 \)&
 \( \frac{g}{4\left( \beta +\mu \right) ^{2}} \)&
 \( \frac{2}{3}\left( 1+\frac{\beta }{\mu }\right)  \)&
\( \frac{\mu }{\mu +\beta } \)\\
 \( b_{b} \)&
 \( -\frac{3}{2\mu } \)&
 \( \frac{3}{2\left| \mu \right| } \)&
 \( \sqrt{1-\frac{9}{2\mu ^{2}}} \)&
 \( \frac{9}{2\mu ^{2}} \)&
 \( \frac{2}{3} \)&
\( 1 \)\\
 \( c_{c} \)&
 \( \frac{2}{3}\beta  \)&
 0 &
 0 &
 \( \frac{4}{9}\beta ^{2} \)&
 \( \frac{6}{4\beta ^{2}+9} \)&
\( 1+\frac{4\beta ^{2}}{9} \)\\
 \( d \)&
 \( -1 \)&
 0 &
 0 &
 1 &
 \( 1/3 \)&
2\\
 \( e \)&
 \( +1 \)&
 0 &
 0 &
 1 &
 \( 1/3 \)&
2\\
 \( f_{b} \)&
 0 &
 0 &
 1 &
 0 &
 \( 2/3 \)&
1\\
\hline
\end{tabular}
\end{center}
\end{table*}

From now on we will discuss the global attractor \( b_{c} \), the
only critical point that may be stationary (i.e. \( \Omega _{c} \)
and \( \Omega _{\phi } \) finite and constant) and accelerated. On
this attractor the two parameters \( \beta  \) and \( \mu  \) are
uniquely fixed by the observed amount of \( \Omega _{c} \) and by
the present acceleration parameter (or equivalently by \( w_{e}=\mu /(\mu +\beta ) \)
). For instance, \( \Omega _{c}=0.20 \) and \( w_{e}=0.23 \) gives
\( \mu =3 \) and \( \beta =10 \). 

Definining the perturbation variables \( \delta _{c}=\delta \rho _{c}/\rho _{c},\quad \delta _{b}=\delta \rho _{b}/\rho _{b},\quad \frac{\sqrt{6}}{\kappa }\varphi =\delta \phi , \)
after algebraic manipulations the following second order equations
for CDM and baryons are found \cite{bias} in the synchronous gauge
for the wavenumber \( k \) :\begin{eqnarray*}
\delta ''_{c}+\frac{1}{2}\left( 4-3w_{e}-4\beta x\right) \delta _{c}'-\frac{3}{2}\gamma \Omega _{c}\delta _{c} & = & 0\\
\delta ''_{b}+\frac{1}{2}\left( 4-3w_{e}\right) \delta _{b}'-\frac{3}{2}\Omega _{c}\delta _{c} & = & 0,
\end{eqnarray*}
where \( H=d\log a/d\tau . \) These equations are valid for subhorizon
scales, that is \( k/H\gg 1 \). Important differences with respect
to the equations of the standard matter-dominated case appear clearly:
the friction term is modified; the constant non-vanishing dynamical
term \( \gamma \Omega _{c} \) \cite{wet95,ame,dam96}, which can
be much larger than unity due to the extra pull of the new interaction,
drives the growth of perturbation even in presence of an accelerated
expansion. The quantities \( x,w_{e} \) and \( \Omega _{c} \) are
given in Tab.1 as functions of the fundamental parameters \( \mu ,\beta  \)
and \( \gamma =1+4\beta ^{2}/3 \). The solutions \cite{bias} are
\( \delta _{c}=a^{m_{\pm }} \)and \( \delta _{b}=ba^{m_{\pm }} \)
where\begin{eqnarray}
m_{\pm } & = & \frac{1}{4}\left[ -4+3w_{e}+4\beta x\pm \Delta \right] \nonumber \\
b_{\pm } & = & 3\Omega _{c}/(3\gamma \Omega _{c}+4\beta xm_{\pm })\nonumber \label{sol} 
\end{eqnarray}
where \( \Delta ^{2}=\left( 24\gamma \Omega _{c}+(-4+3w_{e}+4\beta x)^{2}\right)  \).
The constant \( b\equiv \delta _{b}/\delta _{c}\equiv b_{+} \) is
the bias factor of the growing solution \( m\equiv m_{+} \). The
scalar field solution is \( \varphi \approx (H_{0}a^{(p-1)/p}/k)^{2}\delta _{c}(\beta \Omega _{c}+mbx) \).
For subhorizon wavelengths \( \varphi  \) (which is proportional
to \( \delta \rho _{\phi }/\rho _{\phi } \) ) is always much smaller
than \( \delta _{c},\delta _{b} \) at the present time (although
it could outgrow the matter perturbations in the future if \( p>1 \)).

The solutions \( m_{\pm },b_{\pm } \) apply to all the critical solutions
of Tab. 1 (for \( \Omega _{b}\neq 0 \) the solution can be further
generalized). It is interesting to observe that for \( \mu ,\beta \gg 1 \)
the growth exponent \( m_{+} \) diverges as \( \mu \beta /(\mu +\beta ) \):
the gravitational instability becomes infinitely strong. Let us now
focus on the stationary attractor \( b_{c} \). Four crucial properties
of the solutions will be relevant for what follows: first, the perturbations
grow (i.e. \( m>0 \)) for all the parameters that make the stationary
attractor stable; second, the baryons are antibiased (i.e. \( b<1 \))
for the parameters that give acceleration; third, in the \( k\gg H \)
limit (and in the linear regime), the bias factor is scale independent
and constant in time; and fourth, the bias is independent of the initial
conditions. 

If one studies the perturbations in the Newtonian gauge it can be
shown that a similiar bias develops for the velocity fluctuations,
\( \theta _{b}=b\theta _{c}\, , \) with \( \theta =ik^{i}\delta u_{i}/a, \)
where \( u_{i} \) is the peculiar velocity. This could bring interesting
observational consequences.

The species-dependent coupling generates a biasing between the baryon
and the dark matter distributions. In contrast, the bias often discussed
in literature concerns the distribution only of the very small fraction
of baryons \cite{persic} clustered in luminous bodies and it appears
to depend on luminosity and type \cite{pea}. A first guess could
be that the bulk of baryons follow the distribution of low-luminosity
objects, since they contain most of the mass \cite{vala}. Very recently
it was found \cite{verde} that in the 2dFGRS catalog the average
galaxy bias is close to unity, while galaxies with \( L=L_{*} \)
are slightly antibiased (\( b=0.92\pm 0.11 \)) and galaxies with
\( L\ll L_{*} \) even more so. Moreover, quite remarkably, Verde
et al. (2001) \cite{verde} detected a scale-independent bias from
13 to 65 \( h^{-1} \)Mpc, scales at which our linear calculations
should hold quite well. Similarly, galaxies from the IRAS-PSCz survey
are also antibiased \cite{feld}: \( b=0.8\pm0 .2 \), a result that
agrees with other estimations \cite{rines}. Inclusion of baryons
belonging to weakly clustered objects like Lyman-\( \alpha  \) clouds
can only lower the total baryon bias \cite{dous}. If anything, therefore,
current estimates indicate \( b<1 \) for the total baryon distribution.
To be conservative, here we consider only very broad limits to \( b \):
since the acceleration requires antibias, we assume \( 0.5<b<1 \). 

In Fig. 1 we show all the various constraints. To summarize, they
are: \emph{a}) the present dark energy density \( 0.6<\Omega _{\phi }<0.8 \);
\emph{b}) the present acceleration (\( w_{e}<2/3 \), implying \( \beta >\mu /2 \));
\emph{c}) the baryon bias \( 1>b>0.5 \). On the stationary attractor
there is a mapping between the fundamental parameters \( \mu ,\, \beta  \)
and the observables \( w_{e},\Omega _{\phi } \), so one can plot
the constraints on either pair of variables. It turns out that these
conditions confine the parameters in the small dark shaded area, corresponding
to \( 0.59<w_{e}<0.67 \) or\[
1.1<\beta <1.4,\quad \quad 2.0<\mu <2.6\, .\]
  Therefore, the
parameters of the stationary attractor are determined to within 20\%
roughly. It is actually remarkable that an allowed region exists at
all. The growth rate \( m \) is approximately 0.5 in this region.
For \( b>0.73 \) the possibility of a stationary accelerated attractor
able to solve the coincidence problem would be ruled out.

\begin{figure}
{\centering \resizebox*{7cm}{!}{\includegraphics{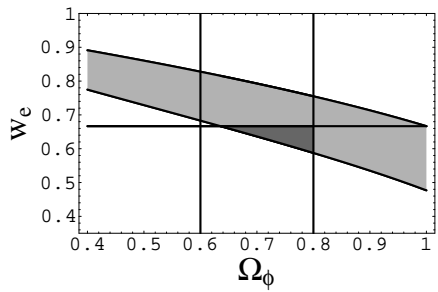}} \par}

\caption{Constraints on the stationary model: below the horizontal line the
expansion is accelerated; in the light grey region the bias is between
0.5 and 1; between the vertical lines \protect\( \Omega _{\phi }\protect \)
is within the observed range. The dark grey region is the surviving
parameter space.}
\end{figure}

The observation of a constant, scale-independent, large-scale antibias
would constitute a strong indication in favor of a dark matter-dark
energy coupling and would indicate a bias mechanism well distinguishable
from the hydrodynamical one. Furthermore the growth rate
\( m \) is an observable quantity that can be employed to test the
stationarity, for instance estimating the evolution of clustering
with redshift. The combined test of \( b \) and \( m \) will be
a very powerful test for the dark matter-dark energy interaction. 

It is reasonable to expect that a similar baryon bias develops whenever
there is a species-dependent coupling; its observation would therefore
constitute a test of the equivalence principle. At the same time,
the species-dependent coupling is requested to provide stationarity
without conflicting with local gravity experiments. Therefore, we
conjecture that the baryon bias is a strong test for all stationary
dynamics.

\end{document}